\documentstyle[amssymb,prb,aps,psfig]{revtex}
\begin{document}

\twocolumn[
\hsize\textwidth\columnwidth\hsize\csname @twocolumnfalse\endcsname

\title{Phonon-mediated thermal conductance of mesoscopic wires with rough edges}
\author{A. Kambili, G. Fagas, Vladimir I. Fal'ko, C.J. Lambert}
\address{School of Physics and Chemistry, Lancaster University, Lancaster
LA1 4YB, UK}
\date{\today}
\maketitle

\begin{abstract}
We present an analysis of acoustic-phonon propagation through long,
free-standing, insulating wires with rough surfaces. Owing to a crossover from 
ballistic propagation of the lowest-frequency phonon mode at
$\omega <\omega _{1}=\pi c/W$ to a diffusive (or
even localized) behavior upon the increase of phonon frequency, 
followed by reentrance into the quasiballistic regime, the heat
conductance of a wire acquires an intermediate tendency to saturate within the
temperature range $T\sim \hbar \omega _{1}/k_{B}$. 
\end{abstract}
]

During recent years, low-temperature heat transport experiments on
electrical insulators have been extended to mesoscopic systems \cite{Roukes}%
, where the wavelength of thermal phonons can be comparable to the
geometrical size of the device. In this regime, phonon transport through a
thermal conductor such as an electrically insulating solid wire, formed from
an undoped semiconductor may exhibit ballistic waveguide propagation \cite
{Blencowe,Kirczenow}. This possibility has stimulated interest in the
guided-wave, phonon-mediated heat conductance, $\varkappa (T)$ of ballistic
wires (with a width $W$ much smaller than the length $L$) connecting a heat
reservoir to a thermal bath \cite{Blencowe,Kirczenow}.

In the present paper, we analyse the low-temperature ($k_{B}T\sim \hbar c\pi
/W$, where c is the sound velocity) heat transport in relatively long ($%
L/W\sim 100$) free-standing insulating wires by taking into account the
effect of surface roughness. The idea behind this analysis is based on the
assumption that, in a long wire, the wire edge or surface roughness may
result in strong scattering, and even in the localisation of acoustic
waves in the intermediate-frequency range, whereas the low-frequency part of
the phonon spectrum would always have ballistic properties due to the
specifics of sound waves. In the high-frequency part of the spectrum,
phonons would have quasiballistic properties, too. This may result in a
non-monotonic temperature dependence of the thermal conductance of such a
system.

To verify the possibility of the existence of such a regime, in principle,
we investigate the dependence on frequency $\omega $ of the transmission
coefficient $\left\langle \Gamma (\omega )\right\rangle $ using a simplified
model of a solid waveguide which has been chosen to reflect two features of
this problem: the influence of roughness on the propagation of vibrations
and the suppression of scattering on the roughness upon the decrease of
the excitation frequency. We approach the problem numerically, by studying
the transmission coefficent averaged over many realizations of a wire
characterized by a given distribution of length scales in the surface
roughness \cite{trans} and using this to find the heat conductance, $%
\varkappa (T)$. Phonon-mediated heat transport in quasi-one-dimensional
systems can be
studied using the same theoretical techniques as electron transport\cite
{Ball,Buttiker}. However, in contrast to electrical conductance, which at
low temperatures is determined by transport properties of electron waves at
the Fermi energy, the thermal conductance of a phonon waveguide is
determined by all phonon energies $\omega $ up to $\hbar \omega \sim k_{B} T$.
This smears out effects of confinement in the transverse direction in the
temperature dependence of $\varkappa (T)$ in ballistic systems \cite
{Blencowe}, but results in a pronounced feature in $\varkappa (T)$ for
strongly disordered free-standing wires. The latter has the form of an
intermediate saturation regime in $\varkappa (T)$ following a linear $T$%
dependence at the lowest temperatures, with an anomalous length dependence
of the saturation value, that scales as $\varkappa _{sat}\propto L^{-1/2}$
for wires with a white-noise spectrum of roughness.

{\it Transmission coefficient analysis and localization of acoustic modes}.
Below, we classify phonon modes in a wire by the number $n$ of nodes in the
displacement amplitude. The $n=0$ lowest-frequency vibrational mode ($\omega
<\omega _{1}=c\pi /W$, where $c$ is the velocity of sound) corresponds to
equal displacements over the cross section of a free-standing wire and has a
linear dispersion; others have frequency gaps, $\omega _{n}(q)=\sqrt{\left(
cq\right) ^{2}+\left( \pi nc/W\right) ^{2}}$. The aforementioned feature
originates from the fact that edge disorder suppresses the transmission of
all coherent phonon modes at high frequencies, but has a little influence on
the $n=0$ mode at $\omega \rightarrow 0$, where $\Gamma (\omega \rightarrow
0)\rightarrow 1$. The transmission coefficient of this mode is mainly
determined by direct backscattering, whose rate depends on the intensity of
the surface roughness harmonic $\left\langle \delta W_{q}^{2}\right\rangle $
with wave number $q\sim \omega /c$. According to Rayleigh scattering theory
in one dimension, the mean free path of this mode diverges at $\omega
\rightarrow 0$ as

\begin{equation}
l_{0}(\omega )\sim \left( \omega _{1}/\omega \right) ^{2}
\left( W^2 / \left\langle \left(
\delta W_{q=\omega /c}\right) ^{2}\right\rangle \right)W,
\end{equation}
even if long-wavelength Fourier components\ $\delta W_{q}$ are equally
represented in the surface roughness $\delta W(x)$. In an infinitely long
wire with white-noise randomness on the surface, this results in a
localization length $L_{\ast }$ for the lowest phonon mode which diverges
at $\omega \rightarrow 0$ as

\begin{equation}
L_{\ast }(\omega )\sim l_{0}(\omega )\propto \omega ^{-2}.  \label{eq:three}
\end{equation}
The latter statement \cite{Fractal} is based on the equivalence between the
localization problem for various types of waves \cite{Elattari}. For a wire
shorter than $L_{\ast }$ scattering yields

\begin{equation}
1-\left\langle \Gamma (\omega )\right\rangle =\alpha \omega ^{2}\;{\rm at}%
\;\omega <\omega _{1}.  \label{lomega}
\end{equation}

In contrast, at higher frequencies, all modes in a wire with a white-noise
randomness backscatter (either via intra- or inter-mode process) typically
at the length scale of $l\sim W/\left\langle \left( \delta W/W\right)
^{2}\right\rangle $. Hence, for frequencies $\omega >\omega _{1}$, the
transmission coefficient tends to follow a linear frequency dependence, $%
\left\langle \Gamma \left( \omega \right) \right\rangle \sim \left( \omega
W/c\right) \left( l/L\right) \ln (W\omega /c)$, which is typical for
diffusion in quasiballistic systems \cite{Pippard,Tesanovich,Leadbeater}.
The crossover from the low-frequency regime to the intermediate-frequency
range in a long enough wire can, therefore, be nonmonotonic, with a
pronounced fall towards zero at $\omega \lesssim \omega _{1}$, similar to
that discussed by Blencowe \cite{Blencoweloc} in relation to the phonon
propagation in thin films. As a result, an irregular wire may exhibit
ballistic phonon propagation at low frequencies, whereas, at higher
frequencies $\omega _{1}\lesssim \omega $, surface roughness would yield diffusive
phonon propagation, or even localization.

The numerical simulations reported below confirm the above naive
expectations. In these simulations, we model the phonons in a crystalline
wire cut from a thin film (with the thickness much less than the wire width)
as longitudinal waves in a two-dimensional strip whose width $W(x)$
fluctuates with rms value $\left\langle \left( \delta W/W\right)
^{2}\right\rangle ^{1/2}=0.1$ on a length scale $\xi $ longer or of order $W$%
. The effect of the width fluctuations consists of the scattering of
acoustic waves propagating along the wire. The model that we adopt here
gives a very simplified representation of a real system, since we ignore the
existence of torsional and transverse bending modes of the wire excitations,
which are known to transfer heat in adiabatic ballistic constrictions \cite{Kirczenow}.
However, it takes into account two features of sound waves: their scattering
and the possible localization by the surface roughness, and the almost ballistic
properties at both ultra-low and high frequencies. In a continuum model,
these lattice vibrations are described by a displacement field $u(x,y)$,
which obeys 2D wave equation inside a wire

\begin{equation}
\omega ^{2}u+c^{2}\nabla ^{2}u=0,  \label{eq:one}
\end{equation}
Displacements obey free boundary conditions ${\bf n}\nabla u=0$ at $y=\pm
(W/2+\delta W_{s}(x))$, where $s=1,2$ indicates the upper and lower edge of
the wire, and {\bf n} stands for the local normal direction to the wire
edge. In the simulations, we discretize Eq. (\ref{eq:one}) on a square
lattice with about 200 sites across the wire\ cross section and, then,
compute $\left\langle \Gamma (\omega )\right\rangle $ numerically using the
transfer matrix method for 100 disorder realisations \cite{RemNum}. Our
numerical code overcomes problems of instability by $QL$ factorising the
transfer matrices at each step \cite{Robinson}. Furthermore, at the end of
each calculation, the S-matrix is checked for unitarity.

Fig. 1 shows the results of such simulations obtained for wires with a
white-noise spectrum of roughness and an aspect ratio $L/W=30$. Both the
low- and high-frequency asymptotic behaviour of transmission coefficient
confirm the expected non-monotonic dependence of $\left\langle \Gamma
(\omega )\right\rangle $ for a wire of length $L\lesssim L_{\ast }$.
Calculations using longer wires (with $L\gg L_{\ast }$) shows a deeper fall
in $\left\langle \Gamma (\omega )\right\rangle $ over a broader range of
frequencies, since phonons with frequencies $\omega \sim \omega _{1}$ behave
as localized vibrations, and their transmission coefficient becomes
exponentialy small. The inset of Fig. 1 shows the corresponding frequency
dependence of the inverse localization length of very long wires (with $%
L/W\sim 1000$), which is approximately in agreement with the result of Eq. (%
\ref{eq:three}). Therefore, the transmission of phonons through rough wires
with white-noise roughness of the edges can be characterised by the
following regimes. At low frequencies, $\omega <\omega _{\ast }$, where $%
\omega _{\ast }(L)\sim \omega _{1}\left( W^{2}/\left\langle \left(%
\delta W_{q\rightarrow 0}\right) ^{2}\right\rangle \right)^{1/2}\sqrt{W/L}$,
$\Gamma \sim 1$ so that
phonons pass through the wire almost ballistically. At intermediate
frequencies, $\omega _{\ast }<\omega \lesssim \omega _{1}$, phonons in the
lowest mode are localized on a length scale shorter than the wire length,
and $\Gamma \rightarrow 0$. Upon a further increase of $\omega $, modes with 
$n\neq 0$ take part in the scattering, the multi-mode localization length
increases, and at $\omega \gtrsim cL/Wl$ the localization length becomes
longer than the sample length, thus restoring a quasi-ballistic character to
phonon transport \cite{Tesanovich,Leadbeater}. The first two regimes of
low-frequency phonon propagation ($\omega <\omega _{1}$) through a wire with 
$L\gg l$ can be jointly described as a function of a single parameter, $%
L/l_{0}(\omega )$ 
\begin{equation}
\left\langle \Gamma (\omega <\omega _{1})\right\rangle =p(\omega /\omega
_{\ast });\;p(0)=1,{\rm \;}p(x\gg 1)\sim e^{-x}.  \label{localized}
\end{equation}
Note that the decline of $\Gamma $ in the vicinity of $\omega \sim \omega
_{1}$ strongly depends on the Fourier spectrum of the roughness. To
illustrate this, we analyzed the effect of roughness composed of harmonics
with wave numbers $q$ restricted to two intervals: (a) $0<q<\pi /W$ and (b)$%
\ \frac{3}{2}\pi /W<q<\frac{7}{2}\pi /W$. The result is shown in Fig. 2 (a)
and (b), respectively. The spectral form of the randomness is relevant,
since it determines the intensity of Bragg-type backscattering processes.
Such processes are the most efficient in forming localization \cite{AltPrig}%
, in which an incident phonon in mode $n$ with wave number $k$ along the
wire axis scatters elastically to mode $n^{\prime }$ with wave number $%
-k^{\prime }$, $k^{\prime }=\sqrt{k^{2}+(n^{2}-n^{\prime 2})(\pi /W)^{2}}$.
Therefore, values of $q=k+k^{\prime }$ represented in the spectrum of $%
\delta W_{q}$ identify the regions of frequencies for which the intra- and
inter-mode Bragg-type scattering is allowed. In Fig. 2, the shaded frequency
intervals indicate the corresponding conditions for two lowest modes, $n=0,1$%
.

{\it Thermal conductance. }In the regime of elastic phonon propagation
through the wire, the heat flow $\dot{Q}$ can be related to the transmission
coefficient of phonons through the wire as

\[
\dot{Q}=\sum_{n,m}\int_{0}^{\infty }\frac{dk}{2\pi }\hbar \omega
_{n}(k)v_{n}(k)(f_{1}(n,k)-f_{2}(n,k))\left| t_{nm}\right| ^{2}, 
\]
where $v_{n}=\frac{d\omega _{n}}{dk}$ is the 1D velocity of a phonon in the
mode $\omega _{n}(k)$, and $f_{1(2)}$ are equilibrium distributions of
phonons at left (right) reservoirs. When the temperature difference $\Delta
T $ between the reservoirs is small \cite{RemTime}, $\Delta T\ll T$, the
thermal conductance, ${\bf \varkappa }=\dot{Q}/\Delta T$, has the form

\begin{equation}
\varkappa =\int_{0}^{\infty }\frac{d\omega }{2\pi }\frac{(\hbar \omega )^{2}%
}{k_{B}T^{2}}\frac{\exp (\hbar \omega /k_{B}T)}{\left[ \exp (\hbar \omega
/k_{B}T)-1\right] ^{2}}\left\langle \Gamma (\omega )\right\rangle  \nonumber
\end{equation}
For a wire with the transmission coefficient shown in Fig. 2(a), where $%
\left\langle \Gamma (\omega )\right\rangle \approx 1+\omega /\omega _{1}$, $%
\varkappa (T)$ is plotted by the dashed-line (1) in Fig. 3, which shows the
crossover from linear to quadratic temperature dependence (at $T\sim
\vartheta _{1}=6\hbar \omega _{1}/k_{B}\pi ^{2}$) discussed in Ref. \cite
{Blencowe}

\begin{equation}
\varkappa \approx \left( k_{B}^{2}\pi /6\hbar \right) T+\left(
0.7k_{B}^{2}/\hbar \right) T^{2}/\vartheta _{1}.  \label{kappaball}
\end{equation}
The ballistic character of heat transport in Eq. (\ref{kappaball}) is
reflected by the independence of $\varkappa $ on the sample length.

In a wire, where the transmission coefficient sufficiently drops at $\omega
\sim \omega _{1}$, as in Fig. 1, we approximate the low-frequency behavior
of the transmission coefficient by a step function, $\left\langle \Gamma
(\omega )\right\rangle =\theta (\omega _{1}-\omega )$, which yields an
intermediate saturation of the thermal conductance at the temperature $T\sim
\vartheta _{1}$,

\begin{equation}
\varkappa (T)\approx \frac{k_{B}\omega _{1}}{2\pi }\left\{ 
\begin{array}{c}
2T/\vartheta _{1},\;T\ll \vartheta _{1}; \\ 
1,\;\vartheta _{1}<T<\vartheta _{1}L/W.
\end{array}
\right.  \label{kappadis}
\end{equation}
The numerical result shown in Fig. 3 by a solid line is in a qualitative
agreement with such an expectation. The horizontal arrow indicates the
saturation value expected from equation (\ref{kappadis}). The upper limit in
the saturation interval mentioned in Eq. (\ref{kappadis}) indicates the
restoration of ballistic conditions for phonon propagation at wavelengths
short enough to avoid wave diffraction at corrugated surfaces \cite
{Tesanovich,Leadbeater}.

Theoretically, the intermediate saturation $\varkappa (T)\approx \varkappa
_{sat}$ at low temperatures is a more robust feature in longer wires of
length $L\gg l\sim W/\left\langle \left( \delta W/W\right) ^{2}\right\rangle 
$, where even the lowest mode, $n=0$ is localized at frequencies $\omega
_{\ast }(L)<\omega <\omega _{1}$. Here $\omega _{\ast }(L)\sim \omega
_{1}\left( W^{2}/\left\langle \left( \delta W_{q\rightarrow 0}\right)
^{2}\right\rangle\right)
^{1/2}\sqrt{W/L}$ is the frequency at which the localization length of $n=0$
acoustic mode is comparable to the wire length. In this case, the saturation
takes place at a lower temperature $\vartheta _{\ast }\sim \hbar \omega
_{\ast }/k_{B}$, and we find that the saturation value of the thermal
conductance within temperature interval $\vartheta _{1}\gtrsim T>\vartheta
_{\ast }$ has an anomalous dependence on the sample length, 
\begin{equation}
\varkappa _{sat}\approx \int_{0}^{\infty }\frac{d\omega }{2\pi }p\left( 
\frac{\omega }{\omega _{\ast }}\right) \sim \frac{k_{B}\omega _{1}}{2\pi }%
\left( \frac{l}{L}\right) ^{1/2}.  \label{lockappa}
\end{equation}
This is an example of a more general scaling law for white-noise roughness;
for a wire with a fractally rough edge, $\left\langle \left( \delta
W_{q}\right) ^{2}\right\rangle \propto q^{z}$, one obtains $\varkappa
_{sat}\propto L^{-1/\left( 2+z\right) }$.

{\it In summary}, our analysis of phonon propagation through long
free-standing insulating wires with rough surfaces has highlighted a feature
in the temperature dependence of the heat conductance $\varkappa (T)$, which
results from the crossover from ballistic propagation of the
lowest-frequency phonon mode at $\omega \ll \omega _{1}$ to diffusive (or
even localized) behavior, with a re-entrance to the quasi-ballistic regime.
Although the model used in this calculation has been restricted to only one
(longitudinal) excitation branch in the wire spectrum, we believe that this
feature persists also in more realistic multi-mode models (which take into
account torsional modes and the wire vibrations of other polarizations),
since all lowest sound modes are scattered by the surface roughness with the
rate decreasing upon the decrease of the frequency. A drastic difference
between phonon transport properties in different frequency intervals results
in a tendency of the heat conductance of a wire to saturate provisionally at
the temperature range of $T\sim hc/Wk_{B}$. An intermediate saturation value
of the wire heat conductance depends on the length of a wire, and, in wires
with length larger than the scattering length of phonons with frequencies $%
\omega \sim \omega _{1}$ has an anomalous length dependence, $\varkappa
_{sat}\propto L^{-1/2}$. 

The authors thank M.Roukes and J.Worlock for attracting our attention to
this problem. This work has been funded in parts by EPSRC and a European
Union TMR programme.

\newpage
\begin{figure}[h]
\centerline{\psfig{figure=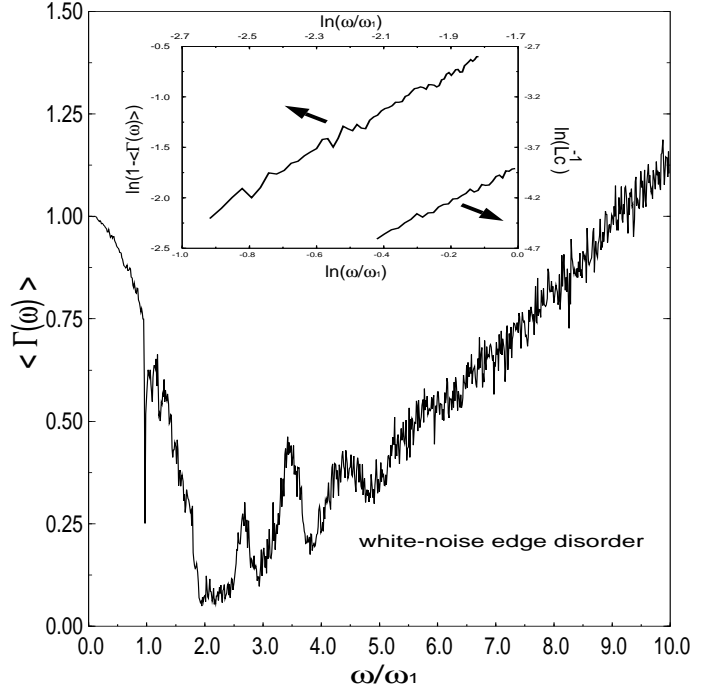,height=10cm,width=10cm}}
\caption{Ensemble-average transmission coefficient for 100 samples versus
frequency for white-noise roughness. Inset: the dependence of $%
(1-\left\langle \Gamma \right\rangle)$ and inverse localization length on
the frequency for $\protect\omega <\protect\omega_{1}$. Linear regression gave
power dependence equal to 1.9 and 1.7, respectively. The arrows in the inset
indicate which pair of axes corresponds to each curve.}
\label{1}
%\newpage
\end{figure}
\begin{figure}[h]
\centerline{\psfig{figure=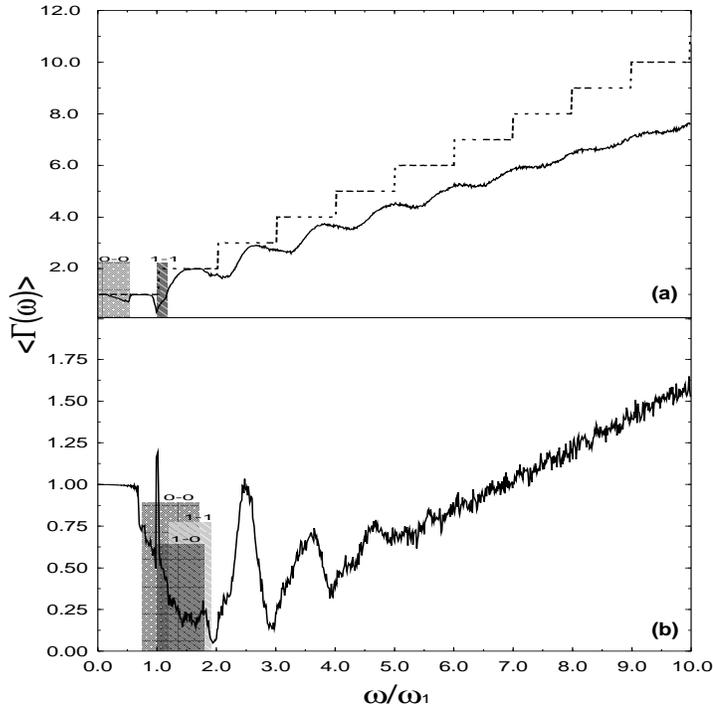,height=10cm,width=10cm}}
\caption{Transmission coefficient versus frequency for two coloured noise
spectra of disorder described in the text. In (a) the dashed line represents the
transmission coefficient for a perfect wire. }
\label{2}
\end{figure}
%\newpage
\begin{figure}[h]
\centerline{\psfig{figure=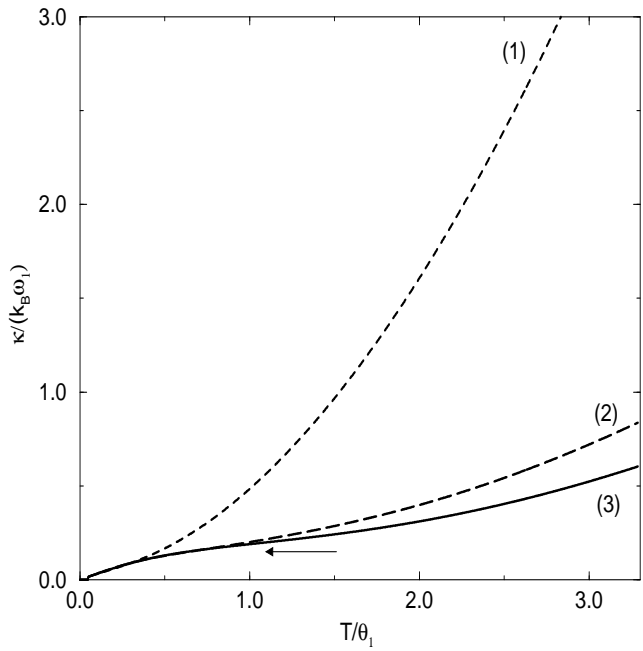,height=10cm,width=10cm}}
\caption{Thermal conductance versus temperature for three different regimes
of disorder. (1) corresponds to the quasi ballistic regime, (2) corresponds
to the disorder regime of figure 2(b), and (3) to the white-noise regime.
The horizontal arrow indicates the saturation value of $\varkappa$. }
\label{3}
\end{figure}

\begin{references}
\bibitem{RemNum}  After discretization, equation of motion takes the form of 
$\varepsilon _{ij}u_{ij}-\gamma \sum_{i^{\prime },j^{\prime
}=n.n.ofi,j}u_{i^{\prime }j^{\prime }}=\omega ^{2}u_{ij}$, where $\gamma
=c^{2}$ and $\varepsilon _{ij}=N_{ij}\gamma $, and $N_{ij}$ is the number of
nearest neighbours of site $j$ from cross section $i$. By representing $%
u_{ij}=\sum^{n_{max}}_{n=0}\Phi _{i}^{n}\chi _{ij}^{n}$, with $\chi
_{ij}^{n}=\sqrt{\frac{2-\delta _{n0}}{W_{i}}}\cos (\frac{n\pi (j-\frac{1}{2})%
}{W_{i}})$ we prepare the equation of motion for the treatment by the
discrete transfer-matrix method, $\sum_{m}I_{nm}^{ii}\Phi _{i}^{m}-\gamma
\sum_{m,\pm }I_{nm}^{ii\pm 1}\Phi _{i\pm 1}^{m}=\omega ^{2}\Phi _{i}^{m}$.
The overlap integrals, $I_{nm}^{ii}=(2\gamma -2\gamma \cos (\frac{n\pi }{%
W_{i}}))\delta _{nm}+\gamma \sum_{j}\sum_{\pm }\beta _{j}^{ii\pm 1}\chi
_{ij}^{m}\chi_{ij}^{n}$, $I_{nm}^{ii\pm 1}=\gamma \sum_{j}\beta _{j}^{ii\pm 1}\chi
_{ij}^{n}\chi _{i\pm 1j}^{m}$ carry information about surface roughness
through the factors $\beta _{j}^{ii^{\prime }}$, which indicate the
existence ($\beta _{j}^{ii^{\prime }}=1$) or absence ($\beta
_{j}^{ii^{\prime }}=0$) of neighbouring sites. The width of each wire
cross-section, $W_{i}=W+\delta W_{1}+\delta W_{2}$ is randomly generated by
introducing a random function $r_{i}^{s}=\sum_{N_{min}}^{N_{max}}a_{m}^{s}%
\sin (\frac{2\pi mi}{L})$, where $N_{min}$ and $N_{max}$ are chosen to yield
the desired harmonic content, $a_{m}$ are randomly taken from the interval $%
[-w/2,w/2]$, and $i$ is the coordinate along the wire axis. Finally, we
normalise the fluctuation of the width to a desired r.m.s. value, so that $%
\delta W_{s}(i)=\sqrt{\left\langle \delta W^{2}\right\rangle /2}\left(
r^{s}(i)-<r^{s}(i)>\right) /\sqrt{<r^{s}(i)^{2}>}$.
\end{references}
\end{document}